\documentclass[fleqn,10pt]{wlscirep}
\RequirePackage{mathtools}

\usepackage{mathrsfs}
\usepackage{amsfonts}
\usepackage{latexsym}
\usepackage{amsthm}
\usepackage{amsmath}
\usepackage{graphicx}
\usepackage{physics}
\usepackage{braket}
\usepackage{xcolor}

\usepackage{CJK}
\usepackage{eucal}

\begin{document}
\begin{CJK*}{UTF8}{gbsn}
\title{The $\mathbb{Z}_2$ toric-code and the double-semion topological order of hardcore Bose-Hubbard-type models in the strong-interaction limit}

\author[1,*]{Wei Wang (王巍)}
\author[2]{Barbara Capogrosso-Sansone}
\affil[1]{Homer L. Dodge Department of Physics and Astronomy, The University of Oklahoma, Norman, Oklahoma 73019, USA}
\affil[2]{Department of Physics, Clark University, Worcester, Massachusetts 01610, USA}
\affil[*]{wei@ou.edu}

\begin{abstract}
We present a generic framework for the emergence of the $\mathbb{Z}_2$ toric-code and the double-semion topological order in a wide class of hardcore Bose-Hubbard-type models governed by density-density interaction and in the strong-interaction regime. We fix fractional filling factor and study under which conditions the density-density interaction gives rise to topological degeneracy. We further specify which dynamics determines the toric-code and the double-semion topological order. Our results indicate that the specifics of the density-density interaction determine the long-range entanglement of the model which possesses ``restricted patterns'' of the long-range entanglement realized in corresponding string-net models with the same topological order.
\end{abstract}


\flushbottom
\maketitle
\end{CJK*}

\section*{Introduction}

Hardcore lattice bosons (HLB) in two spatial dimensions form a wide class of strongly correlated many-body systems which includes models  realizable experimentally with cold atoms and molecules trapped in optical lattices. 
In the past two decades, exotic quantum phases of HLB have attracted a great deal of attention
~\cite{hlb2,hlb3,hlb4,hlb6,entanglementbosehubbard1,hlb7,hlb8,hlb9}. 
Prominent examples include spin liquids and topologically ordered phases in the strong-interaction limit.
In the strong-interaction and strongly-correlated regimes, entanglement can play a prominent role~\cite{entanglementmanybody} making these limits particularly interesting. The quest of HLB models exhibiting certain exotic phases and which can be realized experimentally~\cite{hlb1} is a top priority in view of their astounding potential applications in quantum information and computation\cite{hlbqiqc1,hlbqiqc2}.

HLB are described by Bose-Hubbard-type models where the correlation is governed by a two-site density-density interaction which is diagonal in the Fock basis and expressed as a sum of local operators, $H_0=\sum_{i,j}V_{ij}n_in_j$. It has been reported that for certain lattice geometries, certain $V_{ij}$'s, and appropriate fixed filling factor, strongly interacting HLB harbor $\mathbb{Z}_2$ topological order and form gapped quantum spin liquid~\cite{hlb3,hlb2,entanglementbosehubbard1,hlb4,hlb6}. These observations prompted considerable interest in exploring strongly interacting lattice models which harbor topological order and have the potential to be experimentally realized~\cite{hlbqiqc1,experimenttopo1,experimenttopo2}. This pursuit calls for a general understanding of how certain long-range entanglement~\cite{localunitarytransformation1,topologicalorderbook1} can be emergent in strongly-interacting lattice systems.
Given that quantum phases stabilized in the strong-interaction limit are determined by $H_0$, it is natural to expect that the specific features of the two-site density-density interaction can give rise to patterns of long-range entanglement.

In this paper, we unveil the connection between the specifics of the density-density interaction of hardcore lattice bosons and both the $\mathbb{Z}_2$ toric-code (TC) topological order~\cite{tc1}, and the double-semion (DS) topological order~\cite{sn1} realized in the strong-interaction limit. To the best of our knowledge, this connection is systematically investigated for the first time. We also conjecture a correspondence between hardcore Bose-Hubbard-type lattice models and string-net models by discussing the long-range-entanglement ``patterns''. That is, given the long-range-entanglement ``pattern'' in ground states of certain string-net models, we argue, under certain conditions, the existence of the {\em {same}} topological order in strong-interaction Bose-Hubbard-type lattice models through the realization of a ``restricted pattern'' of long-range-entanglement.
Moreover, with the present work, we expect to provide an analytical guidance for future numerical studies which can pave the way for searching experimentally realizable HLB models harboring spin liquid and topological order.

\section*{Local constraints}
Unlike the exactly solvable Hamiltonians of the TC topological order~\cite{tc1,sn1,tc2} and the DS topological order~\cite{sn1,doublesemion2} containing mutually commutative terms involving three or more sites, in a typical HLB system the interaction term and dynamics terms are non-commutative. Thus, the relationship between the interaction and the topological order harbored by certain HLB models is not as transparent as in exactly solvable models. This relationship can be unveiled by considering the nature of the interaction term. Specifically, the geometry of the lattice, the weight $V_{ij}$, and the filling factor specify certain ``local constraints'', e.g., requirements on the number of particles allowed in a plaquette and their relative position. These ``local constraints" determine the Fock states spanning the ground-state subspace $\mathcal{H}_0$ of ${H}_0$. Moreover, they specify which dynamics leaves $\mathcal{H}_0$ invariant, or equivalently, stipulate how the position of bosons in a given Fock state in $\mathcal{H}_0$ can be moved generating another Fock state in $\mathcal{H}_0$.
As it will become clear in the following discussion, the ``local constraints'' 
determine the capability of strongly-interacting HLB to harbor TC topological order and DS topological order. Throughout this paper, by topologically ordered phase in the strong-interaction limit, we mean a gapped phase (with a spectral gap and a finite ground-state degeneracy), 
extending upon approaching the limit of no dynamics, which has nontrivial bulk topological degeneracy and locally indistinguishable ground states.

In the following, we consider a class of HLB models and study the ``local constraints'' (associated with $H_0=\sum_{i,j}V_{ij}n_in_j$ at specific fractional filling factors) which can give rise to TC or DS topological order in the strong-interaction limit.  
We will first discuss the conditions on the ``local constraints'' which imply the $2^{2g}$-fold topological degeneracy when the underlying lattice has the geometry of a closed orientable surface with genus $g$. Then, we show that certain dynamics determine the TC and the DS topological orders respectively. In summary, we establish a generic framework for the emergence of the TC and the DS topological order in strongly-interacting HLB models.
 Within the framework we have developed, the lattice can possess defects and irregularities, making the system more realistic.



\section*{Model}
The general model describing HLB takes the form 
\begin{equation}
\label{hamiltonian1}H=-t\sum_{(i,j)}a_i^{\dagger}a_j+\sum_{[i,j]}V_{ij}n_i n_j+O(t^2/V),
\end{equation}
where the creation and annihilation operators at site $i$ $a_i^{\dagger}$ and $a_i$ satisfy the hard-core constraint $a_i^{\dagger}a_i^{\dagger}=0$. The first term is the hopping between site $i$ and site $j$, the second term is the diagonal density-density interaction with $n_i=a_i^{\dagger}a_i$, and
the last term represents higher order dynamics terms ($V$ is the maximum of $\abs{V_{ij}}$) which will be discussed below.
We only consider local Hamiltonians. Notice that the hopping and the diagonal interaction can be defined on different, and not-necessarily nearest-neighboring, pairs of sites. In the following we assume fixed fractional filling factor.

\begin{figure}
\centering
\includegraphics[width=0.8\textwidth]{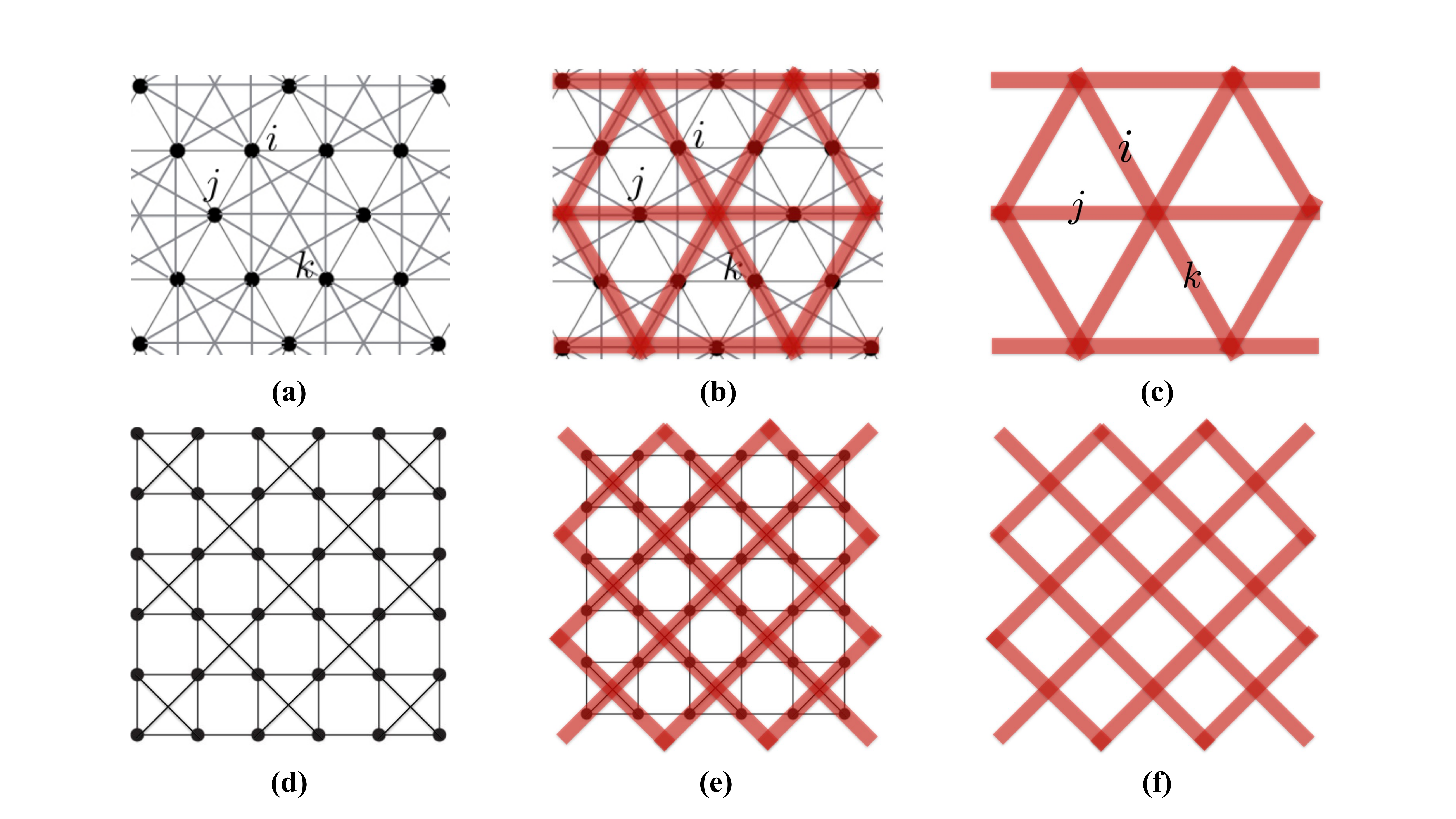}

\caption{\label{figone}(\textbf{b}) shows how to define the modified kagome lattice (\textbf{a}) from the triangular lattice (\textbf{c}). A pair of links (thick red) sharing a vertex in (\textbf{c}) (e.g.,$(i,j)$,$(i,k)$ and $(j,k)$) is replaced by a pair of vertices forming a bond in (\textbf{a}). Similarly, (\textbf{e}) shows how to define the checkerboard lattice (\textbf{d}) from the square lattice (\textbf{f}).}
\end{figure}

For convenience in the following discussion, we can alternatively define $H$ on a lattice $\Lambda$ different from the original lattice $\Lambda_0$. The degrees of freedom, defined on vertices of $\Lambda_0$, are now defined on links of $\Lambda$ as it is done in the toric code model. Note that, $\Lambda$ is not the dual lattice of $\Lambda_0$. In the construction of the dual lattice of $\Lambda_0$, vertices are replaced by plaquettes. Here, vertices are replaced by links. There are multiple ways to define $\Lambda$ from $\Lambda_0$. In many cases of interest, the relation between the two is best elucidated by viewing $\Lambda_0$ as defined from $\Lambda$ in the following manner. The vertex (or site) of $\Lambda_0$ replaces the link (or bond) of $\Lambda$, the pair of sites $(i,j)$ in $\Lambda_0$ form a bond if and only if $(i,j)$ as a pair of links in $\Lambda$ share a vertex, that is, $i$ and $j$ are adjacent links. For example, the modified kagome lattice $\Lambda_0$ (see Fig.~\ref{figone}(a) where bonds are added to the second and third nearest neighbors inside each hexagon of the kagome lattice) is defined from the triangular lattice $\Lambda$ (see Fig.~\ref{figone}(c)). As shown in Fig.~\ref{figone}(b), the sites $i$, $j$ and $k$ in Fig.~\ref{figone}(a) replace the links $i$, $j$ and $k$ in Fig.~\ref{figone}(c). Since each pair of links $(i,j)$, $(i,k)$ and $(j,k)$ share a vertex in Fig.~\ref{figone}(c), pairs of sites $(i,j)$, $(i,k)$ and $(j,k)$ all form bonds in Fig.~\ref{figone}(a). Similarly, as shown in Fig.~\ref{figone}(e), the checkerboard lattice $\Lambda_0$ (see Fig.~\ref{figone}(d)) can be defined from the square lattice $\Lambda$ (see Fig.~\ref{figone}(f)). In the following text, we only consider cases in which $\Lambda$ is \emph{non-bipartite}. This choice will be motivated below, in the discussion of topological degeneracy. It should be noted that $\Lambda_0$ being non-bipartite does not imply that $\Lambda$ is non-bipartite (see Fig.~\ref{figone}(d) and \ref{figone}(f)).

When $t\ll V$, the dynamics of $H$ can be treated perturbatively and the low energy physics of $H$ can be effectively described by a Hamiltonian $H_{\mathrm{eff}}$ solely defined within $\mathcal{H}_0$, the ground-state subspace of $H_0=\sum_{[i,j]}V_{ij}n_in_j$~\cite{hlb3,anyon1}. $H_{\mathrm{eff}}$ consists of the lowest-order perturbative term of the hopping along with terms in $O(t^2/V)$ which leave $\mathcal{H}_0$ unchanged. The detailed form of $H_{\mathrm{eff}}$ depends on $\mathcal{H}_0$ and thus on the ``local constraints'' given by $H_0$. As we shall discuss in details, this effective description reveals the origin of the entanglement in the ground states of $H$: the dynamics, responsible for quantum fluctuation, ``stacks'' Fock states into a single ground state, generating entanglement, while the two-site density-density interaction provides constraints on the contributing Fock states, i.e. they all belong to $\mathcal{H}_0$. 

\section*{Topological degeneracy and anyonic excitation}
We now discuss under which conditions $H_{\mathrm{eff}}$ supports topological degeneracy. In order to do so, we take the following route. In the section Topological sectors, we show that, based on certain requirements in the ``local constraints'', $\mathcal{H}_0$ can be decomposed into $2^{2g}$ topological sectors when the lattice $\Lambda$ has the geometry of a closed orientable surface $\mathcal{S}$ with genus $g$. In the section Effective Hamiltonian, we show that these sectors are invariant under the action of $H_{\mathrm{eff}}$ and any local operators. In the section Ergodicity, we discuss requirements in the ``local constraints''  and the detailed form of $H_{\mathrm{eff}}$ which imply ergodicity of $H_{\mathrm{eff}}$ in each topological sector. Finally, in the sections Toric-code topological order and Double-semion topological order, based on the invariance property and ergodicity, we discuss the presence of topological degeneracy and anyonic excitations corresponding to the TC topological order and the DS topological order respectively.

%


Unless otherwise specified, in the following, we consider the model to be defined on lattice $\Lambda$ where bosons are located on \emph{links}. When necessary, we will use spin language assuming the mapping from occupation number $1$ to spin down $\downarrow$, and from occupation number $0$ to spin up $\uparrow$.

\subsection*{\label{topologicalsectors}Topological sectors}
Given an arbitrary Fock state, we define $s(v)$ as the number of links with occupation number $1$ and adjacent to the vertex $v$, and $P(v)$ (with values in $\{odd, even\}$) as the parity of $s(v)$. In order to build topological sectors of $\mathcal{H}_0$, the ``local constraints'' associated to $H_0$ and the filling factor need to require that the parity $P(v)$ assigned to each vertex $v$ is the same for all Fock states in $\mathcal{H}_0$. We call this requirement the fixed-parity condition. Fixed $P(v)$ at each vertex $v$ for all Fock states in $\mathcal{H}_0$ means that for any two Fock states in $\mathcal{H}_0$, say $\ket{\psi}$ and $\ket{\phi}$, $P(v)_\psi=P(v)_\phi$ for all vertices $v$. Notice that $P(v)$ can be different from vertex to vertex, i.e. $P(v)_\psi\ne P(v')_\psi$ and $P(v)_\phi\ne P(v')_\phi$ for two vertices $v$ and $v'$ while $P(v)_\psi=P(v)_\phi$ and $P(v')_\psi=P(v')_\phi$. For convenience in the following discussion, we denote the set of all Fock states spanning $\mathcal{H}_0$ by $\mathcal{C}_0$. 

\begin{figure}
\centering
\includegraphics[width=0.8\textwidth]{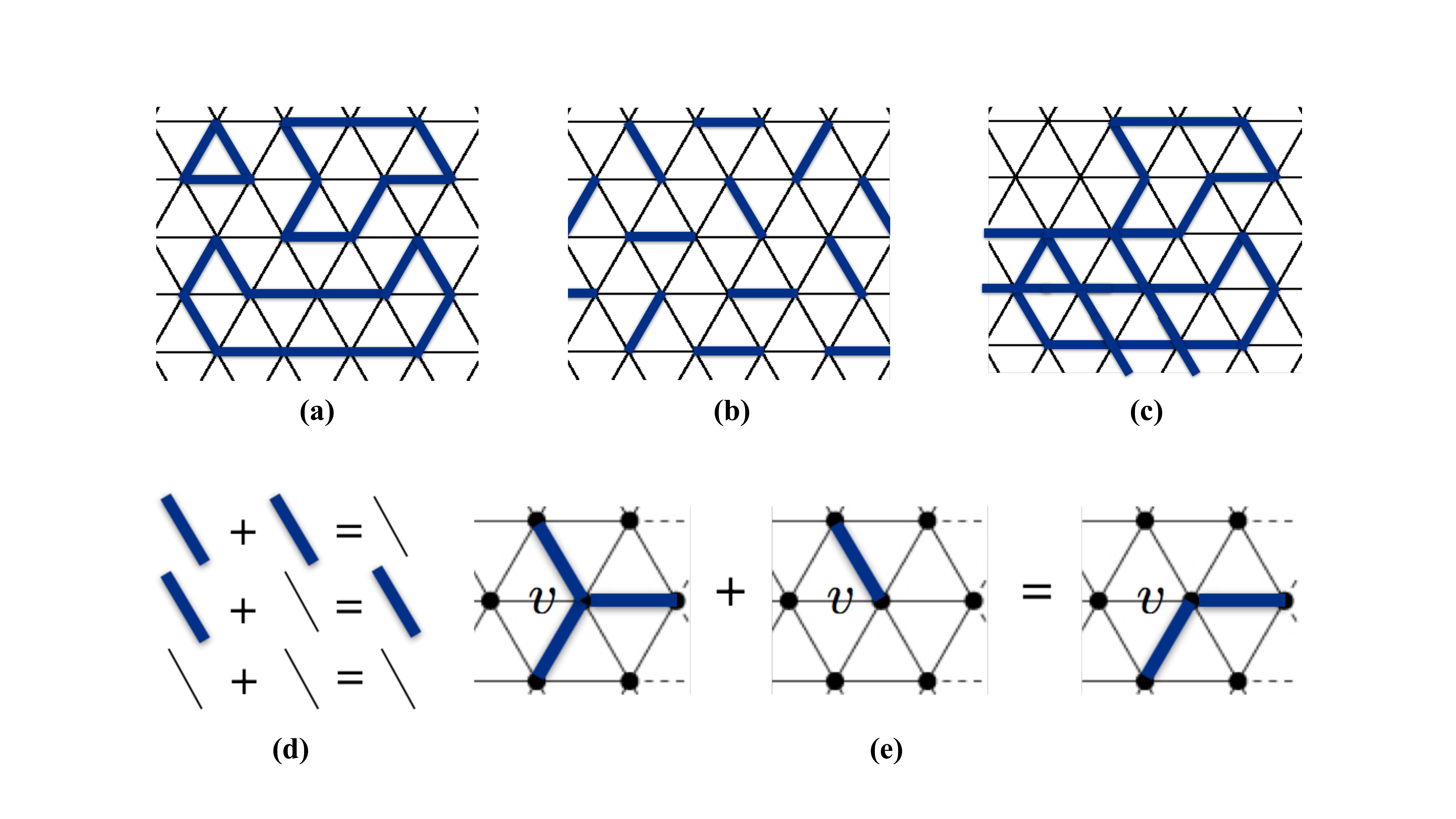}

\caption{\label{figtwo}(\textbf{a}) represents a Fock state belonging to $\mathcal{H}_0$ and corresponding to $1/3$ filling on the triangular lattice. Dark blue links are occupied by bosons. (\textbf{b}) represents a Fock states corresponding to $1/6$ filling. (\textbf{c}) is an example of a Fock state in $\mathcal{C}$ but not in $\mathcal{C}_0$ corresponding to $1/3$ filling. (\textbf{d}) and (\textbf{e}) illustrate the formal summation defined on Fock states.}
\end{figure}

An example of $H_0$ satisfying the fixed-parity condition is $H_0=V\sum_{i,j}n_in_j$ where each $(i,j)$ is a pair of links on the triangular lattice $\Lambda$, which share a vertex (see $(i,j)$ in Fig.~\ref{figone}(c)). Equivalently, each $(i,j)$ can be viewed as a pair of sites forming a bond of the lattice $\Lambda_0$ as shown in Fig.~\ref{figone}(a). It has been shown~\cite{hlb4} that at $1/3$ filling, $s(v)=2$ (and $P(v)=even$) at each vertex. In other words, in each Fock state belonging to $\mathcal{C}_0$, links occupied by bosons form non-crossing closed loops on the lattice (see Fig.~\ref{figtwo}(a) where links in dark blue are occupied by bosons). At $1/6$ filling, $s(v)=1$ (and $P(v)=odd$) at each vertex (see Fig.~\ref{figtwo}(b)), that is,  occupied links in each Fock state in $\mathcal{C}_0$ form a dimer configuration. More generally, $H_0=V\sum_{i,j}n_in_j$ as just defined for the triangular lattice at fixed filling factor implies  $s(v)=const$ also on a generic lattice. Note that fixing only the parity $P(v)$, rather than the number $s(v)$ of occupied links adjacent to a vertex, is a looser condition in the ``local constraints'' which guarantees the existence of topological sectors also for irregular lattices which may have defects or irregularities~\cite{Penna}. 

It should be noted that there exist Fock states satisfying the fixed-parity condition but not consistent with the filling factor. In the case of $1/3$ filling, an example of a Fock state satisfying the fixed parity condition but which does not belong to $\mathcal{C}_0$ is illustrated in Fig.~\ref{figtwo}(c). In the following, in building topological sectors, we denote  by $\mathcal{C}$ the set of all Fock states satisfying the fixed-parity condition but not necessarily consistent with the fixed filling factor. Then, for $H_0$ whose ``local constraints'' obey the fixed-parity condition, we have $\mathcal{C}_0\subset\mathcal{C}$.


We now define a formal summation on Fock states. Based on this summation, topological sectors can be built from the corresponding group-theory structure. We adopt the notation $\star$ for the formal summation to distinguish it from the superposition of states. The formal summation on Fock states is based on the sum of occupation numbers modulo $2$: $1\star1=0$, $1\star0=0\star1=1$, and $0\star0=0$ as illustrated in Fig.~\ref{figtwo}(d) and \ref{figtwo}(e), which is also the group operation of the abelian group $\mathbb{Z}_2=\{0,1\}$. For example, we have $\ket{1,0,0,1\cdots}\star\ket{1,1,0,0\cdots}=\ket{1\star1,0\star1,0\star0,1\star0\cdots}=\ket{0,1,0,1\cdots}$. This rule generalizes the sum rule defining transition graphs in quantum dimer models~\cite{quantumdimer1,quantumdimer2,quantumdimer3}. Based on the above rule, Fock states form an abelian group. In particular, we can define a subgroup $\mathcal{Z}$ in which every Fock state satisfies the fixed-parity condition with $P(v)=even$ at $all$ vertices (e.g., states represented by Fig.~\ref{figtwo}(a) and \ref{figtwo}(c)). Moreover, since in any state with  $P(v)=even$, links occupied by 1 boson form closed loops, also called cycles, on the surface $\mathcal{S}$ on which the lattice $\Lambda$ is embedded, then, we say that $\mathcal{Z}$ consists of Fock states with loop configurations. Loop configurations can be classified according to the topology of $\mathcal{S}$. That is, for a surface $\mathcal{S}$ of genius $g$, $\mathcal{Z}$  can be partitioned into $2^{2g}$ disjoint subsets $\{\mathcal{Z}^1,\cdots,\mathcal{Z}^{2^{2g}}\}$.  Loop configurations belonging to different subsets cannot be connected by local moves such as continuous deformation of loops, or by creation or deletion of loops which delimit a region on the surface $\mathcal{S}$~\cite{homology1}. In order to define topological sectors in $\mathcal{C}$, we partition it in terms of $\{\mathcal{Z}^1,\cdots,\mathcal{Z}^{2^{2g}}\}$ by mapping each state in $\mathcal{C}$ to an even-parity state, i.e. a state in $\mathcal Z$. A simple way to do so is by fixing a reference state belonging to $\mathcal C$ and using the formal summation defined above. Indeed, for an arbitrary reference state $\ket{c_r}=\ket{{c_r}_1,{c_r}_2,\cdots}$ in  $\mathcal{C}$, we can define the map $\ket{c}\mapsto\ket{c}\star\ket{c_r}$ for every $\ket{c}=\ket{c_1,c_2,\cdots}$ in $\mathcal{C}$. Since all elements in $\mathcal{C}$ have the same parity by definition, $\ket{c}\star\ket{c_r}=\ket{c_1\star {c_r}_1,c_2\star {c_r}_2,\cdots}$ is guaranteed to have $even$ parity and hence belongs to $\mathcal{Z}$ as desired. Indeed, $\ket{c}\mapsto\ket{c}\star\ket{c_r}$ maps $\mathcal{C}$ one-to-one onto $\mathcal{Z}$ and defines a partition $\{\mathcal{C}^1,\cdots,\mathcal{C}^{2^{2g}}\}$ of $\mathcal{C}$ such that $\ket{c}$ belongs to $\mathcal{C}^k$ if and only if $\ket{c}\star\ket{c_r}$ belongs to $\mathcal{Z}^k$. Note that for different reference states $\ket{c_r}$, the map $\ket{c}\mapsto\ket{c}\star\ket{c_r}$ is different, but the resultant disjoint subsets $\{\mathcal{C}^1,\cdots,\mathcal{C}^{2^{2g}}\}$ of $\mathcal{C}$ are the same. (See Methods section for a formal discussion of the map.) If we now take the intersection $\mathcal{C}_0\cap\mathcal{C}^k$, then the set $\mathcal{C}_0$ will be partitioned into disjoint subsets $\{\mathcal{C}_0^1,\cdots,\mathcal{C}_0^{2^{2g}}\}$. These subsets span a set of orthogonal subspaces $\{\mathcal{H}_0^1,\cdots,\mathcal{H}_0^{2^{2g}}\}$ which decompose $\mathcal{H}_0$. By construction, these subspaces depend on the topology of the surface $\mathcal{S}$, and as such, they are topological sectors.

\subsection*{\label{effectivehamiltonian}Effective Hamiltonian}
According to the fixed-parity condition, we specify $H_{\mathrm{eff}}$ and show that $H_{\mathrm{eff}}$ does not allow transitions between states belonging to different sectors. Because of the fixed-parity condition associated with $H_0$, the non-vanishing perturbative hopping term of the order $t^2/V$ in $H_{\mathrm{eff}}$ must have the ring-exchange form and be defined on $4$-link closed loops. Likewise, other terms $O(t^2/V)$ which leave $\mathcal{H}_0$ unchanged should have the form of the ring-exchange operators, $a^\dagger_ia_ja^\dagger_ka_l\cdots+\mathrm{H.c.}$. Hence, the effective Hamiltonian has the following generic form,
\begin{equation}
\label{effective1}H_{\mathrm{eff}}=-t^2/V\sum_{b=\{i,j,k,l,\cdots\}}q(b)(a^\dagger_ia_ja^\dagger_ka_l\cdots+\mathrm{H.c.}),
\end{equation}
where $b$ is a closed loop of finite size in the thermodynamic limit, consisting of even number of adjacent links $\{i,j,k,l,\cdots\}$ in $\Lambda$, {\emph{contractible}}, i.e. delimiting a region, on $\mathcal{S}$.
The value of the coefficient $q(b)=\pm 1$ will be specified below. For each $b=\{i,j,k,l,\cdots\}$ contributing to the sum in Eq.~\ref{effective1}, we can equivalently express the operators $a^\dagger_ia_ja^\dagger_ka_l\cdots+\mathrm{H.c.}$ as $\dyad*{\bar{b}}{\underbar{b}}+\dyad*{\underbar{b}}{\bar{b}}$, where  $\ket{\bar{b}}$=$\ket{1,0,1,0,\cdots}$ and $\ket{\underbar{b}}$=$\ket{0,1,0,1,\cdots}$ stand for states solely defined on links forming the loop $b$ with bosons located alternatively on the loop $b$. Alternatively, we can write the ring-exchange operator for each $b$ using Pauli operators as $\Gamma_{\mathcal{H}_0}\Gamma_b\prod_{i\in b}\sigma^x_i$ where $\Gamma_{\mathcal{H}_0}$ is the projection operator associated with $\mathcal{H}_0$ and $\Gamma_b=\dyad*{\bar{b}}{\bar{b}}+\dyad*{\underbar{b}}{\underbar{b}}$ is the projection operator defined on the closed loop $b$.

We now show that an arbitrary subspace $\mathcal{H}_0^k$ is invariant under the action of $H_{\mathrm{eff}}$ by using the Pauli operator expression. Let us consider a Fock state $\ket{c}=\ket{c_1,c_2,\cdots}$ in $\mathcal{C}_0^k$. Then, as discussed above, 
given a reference state $\ket{c_r}$ in $\mathcal{C}$ and the map $\ket{c}\mapsto\ket{c}\star\ket{c_r}$ which is used to define the partition $\{\mathcal{C}^1,\cdots,\mathcal{C}^{2^{2g}}\}$, $\ket{c}\star\ket{c_r}=\ket{z}$ belongs to $\mathcal{Z}^k$.
When acting with $\Gamma_{\mathcal{H}_0}\Gamma_b\prod_{i\in b}\sigma^x_i$ upon $\ket{c}$, the result is $0$ if $\ket{c}$ does not have bosons occupying alternating links on the loop $b$, or $\ket{c}\star\ket{b}$ where $\ket{b}$ is a Fock state with occupation number $b_i=1$ for each link $i$ on loop $b$ and $b_i=0$ otherwise. Moreover, $\ket{c}\star\ket{b}\star\ket{c_r}=\ket{c}\star\ket{c_r}\star\ket{b}=\ket{z}\star\ket{b}$, where $\ket{z}\star\ket{b}$ can be obtained from $\ket{z}$ by acting upon with local moves. Hence, $\ket{z}\star\ket{b}$ belongs to $\mathcal{Z}^k$. Consequently, $\ket{c}$ and $\ket{c}\star\ket{b}$ are mapped to the same subspace $\mathcal{Z}^k$, that is, they both belong to $\mathcal{C}_0^k$. Therefore, $\Gamma_{\mathcal{H}_0}\Gamma_b\prod_{i\in b}\sigma^x_i$ maps $\mathcal{H}_0^k$ into itself, that is, $\mathcal{H}_0^k$ is invariant under the action of $H_{\mathrm{eff}}$. Moreover, since an arbitrary local operator is expressed as a sum of products of Pauli operators and can be expressed in terms of local moves, it is easy to show that $\mathcal{H}_0^k$ is invariant under the action of any local operator.

\begin{figure}
\centering
\includegraphics[width=0.85\textwidth]{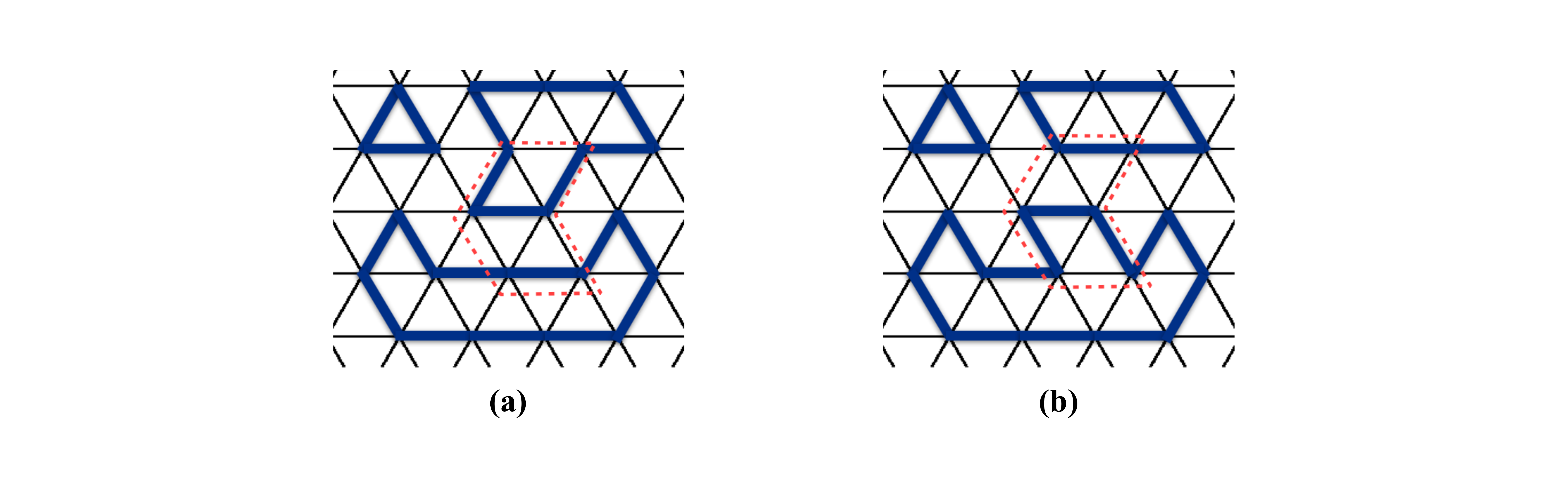}

\caption{\label{figthree}(\textbf{a}) and (\textbf{b}) represent two Fock states in $\mathcal{C}_0$ corresponding to $H_0=V\sum_{i,j}n_in_j$, where the sum extends to all pairs of adjacent links defined on  triangular lattice and at $1/3$ filling. (\textbf{b}) is obtained from (\textbf{a}) by exchanging bosons (colored links) and holes (uncolored links) on a closed loop $b'$ indicated by the red dashed line.}
\end{figure}

\subsection*{\label{ergodicity}Ergodicity}
The irreducibility of $\mathcal{H}_0^k$ under the action of $H_{\mathrm{eff}}$ is equivalent to the ergodicity of $H_{\mathrm{eff}}$ in each $\mathcal{H}_0^k$. Here, ergodicity means that each $\mathcal{C}_0^k$ is nonempty and any two Fock states in $\mathcal{C}_0^k$ can be connected by sequentially applying ring-exchange operators in $H_{\mathrm{eff}}$.  Rigorously proving that ergodicity holds is very challenging as efforts towards this task for the case of quantum dimer model ($s(v)=1$) have shown~\cite{quantumdimer1}. This is beyond the scope of the present work. We therefore take a heuristic approach and discuss ergodicity in terms of the two requirements discussed below. The heuristic arguments used here can be viewed as a generalization of the conditions which guarantee ergodicity of quantum dimer models defined on non-bipartite lattices~\cite{quantumdimer3,quantumdimer2}. We should notice that, in principle, ergodicity can be numerically checked by Monte Carlo simulations~\cite{hlb2}.


First, ``local constraints'' need to require that given a Fock state $\ket{c}$ in $\mathcal{C}_0$, for any closed loop $b'$ consisting of even number of links --but not necessarily contributing to the sum of $H_{\mathrm{eff}}$, hence not necessarily contractible-- and such that bosons are located alternatively on it,
positions of bosons and holes on $b'$ can be exchanged resulting in another Fock state $\ket{c'}$ in $\mathcal{C}_0$. This is illustrated in Fig.~\ref{figthree}(a) (state $\ket{c}$) and Fig.~\ref{figthree}(b) (state $\ket{c'}$) where the loop $b'$ is indicated by the red dashed line. It is important to notice that this exchange between bosons and holes does not change $s(v)$. Moreoever, if $b'$ is contractible, states $\ket{c}$ and $\ket{c'}$ necessarily belong to the same sector since the boson-hole exchange corresponds to a local operation. On the other hand, if $b'$ is non-contractible, $\ket{c}$ and $\ket{c'}$ belong to different sectors. The existence of non-contractible $b'$ loops (e.g., loops in the two toroidal directions on a torus) is guaranteed by the fractional filling factor. The requirement just described associated to ``local constraints'' ensures that each sector is non-trivial since the existence of state $\ket{c}$ implies the existence of a state $\ket{c'}$ in each sector.

Secondly, for any state $\ket{c}$ in $\mathcal{C}_0$, the ``local constraints'' need to imply that the surface $\mathcal{S}$ on which the lattice is embedded can be covered by areas $\mathcal{A}$, in general partially overlapping, bounded by contractible closed loops $b'$ on which bosons are located alternatively. Notice that the areas should be of finite-size. This requirement together with the first one ensures that no state in $\mathcal{C}_0^k$ is isolated from others upon applying ring-exchange operators. Moreover, since Eq.~\ref{effective1} can be viewed as a generalization of the dynamics in quantum dimer models, we also require $\Lambda$ to be non-bipartite. If the $b$ loops entering the sum in Eq.~\ref{effective1} satisfy the two-requirements described above and $\Lambda$ is non-bipartite, then, based on this heuristic argument, we conclude that the Hamiltonian in Eq.~\ref{effective1} is ergodic in each $\mathcal{C}_0^k$. 

 Examples of systems whose ``local constraints'' are consistent with the requirements discussed above include models with $H_0=V\sum_{i,j}n_in_j$ where each $(i,j)$ is a pair of links sharing a vertex on lattice $\Lambda$. See references~\cite{hlb2,hlb3,hlb4,hlb6} and Equation (12) in~\cite{doublesemion2}. It is an interesting and open question whether and under which conditions the topologically ordered phase corresponding to uniform weights $V_{i,j}=V$ survives when this constraint is relaxed. This question can possibly be answered using quantum Monte Carlo simulations to calculate topological entanglement entropy, and we are currently working on it.

\subsection*{\label{tc}Toric-code topological order}
In order to demonstrate topological degeneracy, we are left with showing that ground states $\ket{\psi_{\mathrm{ex}}^k}$ of $H_{\mathrm{eff}}$ in each $\mathcal{H}_0^k$ are unique, have the same energy which is gapped above, and are locally indistinguishable.

We first study the case of the TC topological order which corresponds to 
\begin{equation}
\label{TC}q(b)=1
\end{equation}
for all $b$ in $H_{\mathrm{eff}}$. Due to the ergodicity of $H_{\mathrm{eff}}$ in $\mathcal{H}_0^k$, the Perron-Frobenius theorem~\cite{Tasaki1} implies the uniqueness of the lowest-energy state $\psi_{\mathrm{eff}}^k$ in $\mathcal{H}_0^k$ in a finite-size system. We now discuss under which conditions ground-states in different sectors are degenerate and separated from other eigenstates by an energy gap independent on the system size. In some special cases, like the quantum dimer model on the Kagome lattice~\cite{quantumdimer2}, $H_{\mathrm{eff}}$ itself is exactly solvable and gapped, and ground-states are equal-weight sums of Fock states in each topological sector. In this case, $H_{\mathrm{eff}}$ has $2^{2g}$-fold topological degeneracy. More generally, for non-exactly solvable effective models, $H_{\mathrm{eff}}$ can be smoothly connected to an exactly solvable model $H_{\mathrm{ex}}$, analogous to the Rokhsar-Kivelson (RK) quantum dimer model~\cite{quantumdimer1}, by means of the $r$-parameterized Hamiltonian
\begin{equation}
\label{rparatc}\tilde H(r)=-t^2/V\sum_b(\dyad*{\bar{b}}{\underbar{b}}+\dyad*{\underbar{b}}{\bar{b}})+r\sum_b\Gamma_b,
\end{equation}
where $b$ is a contractible loop consisting of an even number of adjacent links in $\Lambda$, $\ket{\bar{b}}$=$\ket{1,0,1,0,\cdots}$ and $\ket{\underbar{b}}$=$\ket{0,1,0,1,\cdots}$ stand for states solely defined on links forming the loop $b$ with bosons located alternatively, and $\Gamma_b=\dyad*{\bar{b}}{\bar{b}}+\dyad*{\underbar{b}}{\underbar{b}}$.
If $r=0$, then $\tilde H=H_\mathrm{eff}$. If $r=t^2/V$, then one obtains an exactly solvable model $\tilde H(r)=H_{\mathrm{ex}}$ with
\begin{equation}
\label{extc}H_{\mathrm{ex}}=t^2/V\sum_b(\ket{\bar{b}}-\ket{\underbar{b}})(\bra{\bar{b}}-\bra{\underbar{b}}). 
\end{equation}
The lowest-energy eigenstate $\ket{\psi_{\mathrm{ex}}^k}$ of $H_{\mathrm{ex}}$ in each $\mathcal{H}_0^k$ is an equal-weight sum of all Fock states in $\mathcal{C}_0^k$ and has zero energy. 
For RK quantum dimer models (corresponding to Eq.~\ref{extc} with $s(v)=1$) on non-bipartite lattices $\Lambda$, it has been shown that there exists an energy gap separating ground states $\ket{\psi_{\mathrm{ex}}^k}$ in each sector from other eigenstates~\cite{quantumdimer1}. This conclusion has also been demonstrated for models (included in our generic models) other than quantum dimer models and corresponding to $s(v)=2,3$~\cite{hlb4,hlb3}. In reference~\cite{quantumdimer1}, the authors indicate that the RK quantum dimer model possesses $2^{2g}$-fold topological degeneracy on a torus, and that the gap properties only depend on the graph nature of the lattice, i.e. bipartite or not, and on the ergodicity of the dynamics in $\mathcal{H}_0$. Given that our models can be viewed as generalizations of quantum dimer models, we believe these results can be extended to models considered here. Therefore, we \emph{expect} $H_{\mathrm{ex}}$ to possess a gap and have a $2^{2g}$-fold topological degeneracy.

Let us call the subspace spanned by ground-states $\{\ket{\psi_{\mathrm{ex}}^k}\}$ of $H_{\mathrm{ex}}$ as $\mathcal{E}_{\mathrm{ex}}$. If the gap remains open as the parameter $r$ is tuned to $0$, then $H_{\mathrm{eff}}$ will also be gapped and have a $2^{2g}$-fold topological degeneracy. Examples of models for which the gap remains open include models corresponding to Eq.~\ref{rparatc} with $s(v)=2$ and $s(v)=3$ on the triangular lattice~\cite{hlb4} and $s(v)=1$ on a kagome lattice or star lattice~\cite{quantumdimer2,doublesemion1}. Furthermore, as indicated in Ref.~\cite{localunitarytransformation1}, local and gapped Hamiltonians smoothly connected by a family of local Hamiltonians which maintain the gap open are in the same quantum phase with their ground state subspace connected by a local unitary transformation. As a result, there exists a local unitary transformation $\mathcal{U}_{\mathrm{ex}}$ determined by $\tilde H(r)$, such that $\mathcal{U}_{\mathrm{ex}}(\mathcal{E}_{\mathrm{eff}})=\mathcal{E}_{\mathrm{ex}}$, where $\mathcal{E}_{\mathrm{eff}}$ is the subspace spanned by $\{\ket{\psi_{\mathrm{eff}}^k}\}$. As a consequence, $H_{\mathrm{eff}}$ and $H_{\mathrm{ex}}$ are in the same quantum phase. Specifically, if the ground states $\{\ket{\psi_{\mathrm{ex}}^k}\}$ are $locally$ $indistinguishable$, then both $H_{\mathrm{ex}}$ and $H_{\mathrm{eff}}$ are in the same topologically ordered phase.

We now discuss local indistinguishability of $\{\ket{\psi_{\mathrm{ex}}^k}\}$ in terms of the ``local constraints''. Local indistinguishability means that $\mel{\psi_{\mathrm{ex}}^k}{A}{\psi_{\mathrm{ex}}^l}=f(A)\delta_{k,l}$ in the thermodynamic limit for any local operator $A$, where $f(A)$ is independent of topological sectors. According to our discussion on topological sectors, a local operator $A$ cannot induce any transition between different topological sectors, and thus we have $\mel{\psi_{\mathrm{ex}}^k}{A}{\psi_{\mathrm{ex}}^l}=0$ for $k\ne l$. Then, we only need to show that the value of $f(A)$ is the same in different topological sectors. Let's consider a local region including the support of $A$ and delimited by a closed loop $b$, and a Fock state $\ket{\mathbf{b}}$ solely defined on this region and satisfying the ``local constraints''. Starting from $\ket{\mathbf{b}}$, one can imagine building Fock states in each topological sector by populating  with particles links outside $b$ according to ``local constraints''. Since  ``local constraints'' stipulate how bosons are located on the lattice \emph{locally}, the procedure of building Fock states is independent of topological sectors which differ from each other by {\em{nonlocal}} properties. As a consequence, one would expect that the number of Fock states in each sector coinciding with $\ket{\mathbf{b}}$ approaches the same value in the thermodynamic limit.

Note that this property can in principle be checked numerically. Finally, we can conclude that averages of $A$ in each degenerate ground state $\ket{\psi_{\mathrm{ex}}^k}$ (which is an equal-weight sum of Fock states) approach the same value in the thermodynamics limit, i.e. $f(A)$ is independent of topological sectors.

Following the discussion above, $H_{\mathrm{eff}}$ and $H_{\mathrm{ex}}$ have the same topological order and the anyons realized by excitations of $H_{\mathrm{eff}}$ are the same as the ones realized in $H_{\mathrm{ex}}$. 
Excitations of $H_{\mathrm{ex}}$ are essentially the same as excitations of a Rokhsar-Kivelson quantum dimer model in the RVB phase~\cite{quantumdimer1,rvb1}. There are three types of anyonic excitations: (1) deconfined spin-$1/2$ fermions (analogous to spinons in quantum dimer model) appearing in pairs, corresponding to two ends of an open string in $\Lambda$, for which a typical excited state is an equal-weight sum of Fock states with two vertices with assigned parities opposite to $P(v)$; (2) spinless bosons (analogous to visons) 
appearing in pairs, corresponding to two ends of an open string in the dual lattice of $\Lambda$, for which the excitation can be viewed as a locally dressed operator of $\prod\sigma^z$ on an open string in the dual lattice~\cite{rvb1}; (3) spinon-vison composition. Fusion and braiding of the three types of anyons can be studied in the same way as in the quantum dimer model of RVB phase~\cite{quantumdimer1,rvb1}. Accordingly, the anyonic excitations defines the $\mathbb{Z}_2$ toric-code topological order~\cite{tc1} of model~\ref{effective1} with $q(b)=1$.

\subsection*{\label{ds}Double-semion topological order}
In order to study the case of DS topological order, we can similarly define
\begin{equation}
\label{HexDS}H_{\mathrm{ex}}^{\mathrm{DS}}=t^2/V\sum_b(\ket{\bar{b}}-q(b)\ket{\underbar{b}})(\bra{\bar{b}}-q(b)\bra{\underbar{b}})
\end{equation}
connected to $H_{\mathrm{eff}}^{\mathrm{DS}}$ via the $r$-parametrized Hamiltonian
\begin{equation}
\label{rparatcds}\tilde{H}^{\mathrm{DS}}(r)=-t^2/V\sum_bq(b)(\dyad*{\bar{b}}{\underbar{b}}+\dyad*{\underbar{b}}{\bar{b}})+r\sum_b\Gamma_b \;\; 
\end{equation}
where $b$, $\ket{\bar{b}}$, $\ket{\underbar{b}}$, and $\Gamma_b$ are defined as in Eq.~\ref{rparatc}.
First, we specify the phase factor $q(b)$ according to considerations below. Then, we show that the DS topological order is implied by the specified phase factor.

As shown in the exactly solvable model of the DS topological order~\cite{sn1,topologicalorderbook1}, the anyonic excitations are determined by the fact that the ground state is a sum of \emph{non-crossing-closed-loop} configurations on a trivalent lattice with expansion coefficients $(-1)^{N(z)}$ where $N(z)$ is the number of loops in $\ket{z}$~\cite{topologicalorderbook1}. (Notice that non-crossing-closed-loop configurations belong to $\mathcal{Z}$ by definition). In order to study the DS topological order of $H_{\mathrm{ex}}^{\mathrm{DS}}$ defined on an arbitrary non-bipartite lattice $\Lambda$, we will ``equivalently'' represent the ground state of $H_{\mathrm{ex}}^{\mathrm{DS}}$ by a sum of non-crossing-closed-loop configurations with expansion coefficients determined by the number of loops. Note that the number of loops can be uniquely defined only for non-crossing-closed-loop configurations. One obvious way to realize non-crossing-closed-loop configurations corresponds to $s(v)=0$ or $2$ for all vertices so that, in each $\ket{c}$ belonging to $\mathcal{C}_0$, links with occupation number $1$ directly form non-crossing closed loops, see e.g. Fig.~\ref{figthree}(a). However, in this case, to realize the expansion coefficients $(-1)^{N(c)}$, the phase factor $q(b)$ must be replaced by an operator, similar to the operator-valued phase factor in the exactly solvable model~\cite{sn1,doublesemion2}. Adopting the operator-valued phase factor will deviate from our goal of making Eq.~\ref{hamiltonian1} as realistic as possible. Hence, we consider another way to realize non-crossing-closed-loop configurations by considering $s(v)=1$ for all vertices. In this case, in order to represent the ground state of $H_{\mathrm{ex}}^{\mathrm{DS}}$ as a sum of non-crossing-closed-loop configurations, we make use of a local unitary transformation, as explained below. 

By fixing a reference Fock state $\ket{c_r}$ in $\mathcal{C}_0$, we can define a quantum circuit operator $\mathcal{U}'=\prod_{{c_r}_i=1}\sigma^x_i$ which is local unitary~\cite{localunitarytransformation1}. Then,  we construct the locally and unitarily transformed Hamiltonian $\mathcal{U}'H_{\mathrm{ex}}^{\mathrm{DS}}\mathcal{U}'^\dagger$ which possesses the same quantum phase as $H_{\mathrm{ex}}^{\mathrm{DS}}$. Let us assume that the ground state $\ket{\psi_{\mathrm{ex}}^k}$ of $H_{\mathrm{ex}}^{\mathrm{DS}}$ is a sum of $\ket{c}$ belonging to $\mathcal{C}_0^k$ (we will later show that this is indeed the case), then, the ground state of $\mathcal{U}'H_{\mathrm{ex}}^{\mathrm{DS}}\mathcal{U}'^\dagger$, $\mathcal{U}'\ket{\psi_{\mathrm{ex}}^k}$, is a sum of $\mathcal{U}'\ket{c}=\ket{c}\star\ket{c_r}$. Due to the fixed parity condition, $\ket{c}\star\ket{c_r}$ belongs to $\mathcal{Z}$ with $s(v)=0,2$ at each vertex and, as a cosequence, has a non-crossing-closed-loop configuration. Thus, $\mathcal{U}'\ket{\psi_{\mathrm{ex}}^k}$ is the desired ``equivalent'' representation of $\ket{\psi_{\mathrm{ex}}^k}$. 


Now, in terms of the fixed reference state $\ket{c_r}$ in $\mathcal{C}_0$, we specify $q(b)$ in $H_{\mathrm{eff}}^{\mathrm{DS}}$ defined on an arbitrary non-bipartite lattice $\Lambda$. In terms of the specified $q(b)$, we will show that the ground state of $H_{\mathrm{ex}}^{\mathrm{DS}}$ is a superposition of $\ket{c}$ belonging to $\mathcal{C}_0^k$ with expansion coefficients depending on the number of loops in $\ket{c}\star\ket{c_r}$. 
\begin{figure}
\centering
\includegraphics[width=0.85\textwidth]{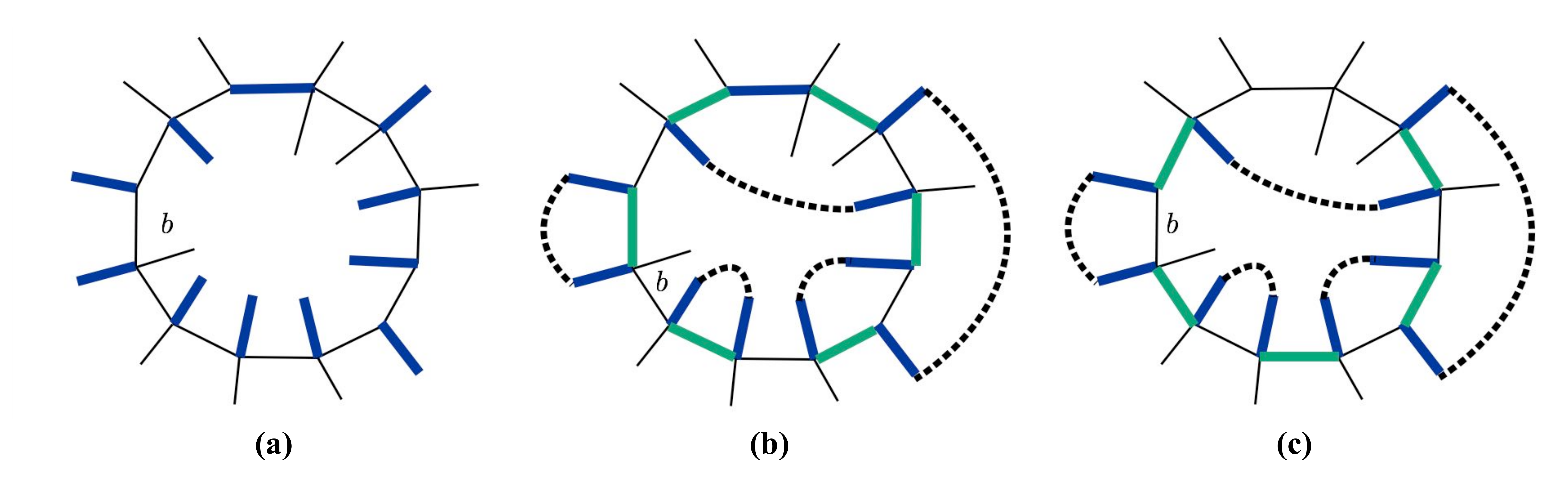}
\caption{\label{figfour}(\textbf{a}) represent an arbitrary loop $b$ contributing to the sum of $H_{\mathrm{eff}}^{\mathrm{DS}}$ and links emanating from it, where blue color represents occupied links in the reference state $\ket{c_r}$, black links are unoccupied. We omit the lattice structure complementary to the loop $b$ and links emanating from $b$. (\textbf{b}) represents $\ket{c}\star\ket{c_r}$ and (\textbf{c}) represents $\ket{c'}\star\ket{c_r}$. Green color represents occupied links in state $\ket{c}$ and $\ket{c'}$, black dashed lines complete the loops outside and inside $b$.}
\end{figure}

Let us consider an arbitrary loop $b$ contributing to the sum of $H_{\mathrm{eff}}^{\mathrm{DS}}$. Fig.~\ref{figfour}(a) shows the loop $b$ and links emanating from it, where blue color represents occupied links in the reference state $\ket{c_r}$ and black links are unoccupied. We omit the lattice structure complementary to the loop $b$ and the links emanating from it. Here, loop $b$ in Fig.~\ref{figfour}(a) is made long enough to illustrate general arguments below while in well known quantum dimer models~\cite{quantumdimer3,doublesemion1} loops contributing to Eq.~\ref{effective1} are shorter. Consider an arbitrary $\ket{c}$ in $\mathcal{C}_0$ in which bosons are located alternatively on $b$. Note that the existence of such a $\ket{c}$ is guaranteed by the way we specify the dynamics in Eq.~\ref{hamiltonian1} as discussed in the Ergodicity section. We want to specify $q(b)$ only in terms of $b$ and $\ket{c_r}$ and such that, given the state $\ket{c'}=(\dyad*{\bar{b}}{\underbar{b}}+\dyad*{\underbar{b}}{\bar{b}})\ket{c}$, when $q(b)=1$ the number of loops in $\ket{c'}\star\ket{c_r}$ has the same parity as in $\ket{c}\star\ket{c_r}$; when $q(b)=-1$ the parity is changed. Since $\dyad*{\bar{b}}{\underbar{b}}+\dyad*{\underbar{b}}{\bar{b}}$ only acts on the loop $b$, the change in the number of loops only depends on how the loops are reconnected on $b$. As an example, Fig.~\ref{figfour}(b) represents $\ket{c}\star\ket{c_r}$ and Fig.~\ref{figfour}(c) represents $\ket{c'}\star\ket{c_r}$, where blue links correspond to occupied links in $\ket{c_r}$ while green links correspond to occupied links in $\ket{c}$ or $\ket{c'}$ respectively, black links are unoccupied, and black dashed lines complete the loops outside and inside $b$. Note that, if a green and a blue link overlap, the link will appear black in the representation of $\ket{c}\star\ket{c_r}$ or $\ket{c'}\star\ket{c_r}$ due to the sum rule $1\star1=0$. The action of $\dyad*{\bar{b}}{\underbar{b}}+\dyad*{\underbar{b}}{\bar{b}}$ corresponds to shifting the green links clockwise (or counterclockwise) on $b$. As we show in the Methods section, $q(b)$ is defined in terms of the number of outer and inner links occupied by bosons in state $\ket{c_r}$ and emanating from vertices on $b$. We will define two even numbers, $M_1(b,c_r)$ and $M_2(b,c_r)$, by reducing the number of outer and inner links emanating from $b$ and occupied by bosons in $\ket{c_r}$, so that the parity change in the number of loops from $\ket{c}\star\ket{c_r}$ to $\ket{c'}\star\ket{c_r}$ can be characterized by the number of removable links $M_1(b,c_r)$ and the number of non-removable links $M_2(b,c_r)$. Accordingly, we set
\begin{equation}
\label{DS}q(b)=i^{M_1(b,c_r)+M_2(b,c_r)+2}.
\end{equation}
As shown in the Methods section such defined $q(b)$ satisfies the desired properties.

We now define state $\ket{\psi_{\mathrm{ex}}^k}=\sum_{c\in\mathcal{C}_0^k}-1^{N(\ket{c}\star\ket{c_r})}\ket{c}$ where $N(\ket{c}\star\ket{c_r})$ is the number of loops in the state $\ket{c}\star\ket{c_r}$ and show that it is a unique ground state in the topological sector $\mathcal{H}_0^k$. Indeed, according to above discussion on the properties of $q(b)$, when $q(b)=1$, $(\ket{\bar{b}}-\ket{\underbar{b}})(\bra{\bar{b}}-\bra{\underbar{b}})\ket{\psi_{\mathrm{ex}}^k}=0$; when $q(b)=-1$, $(\ket{\bar{b}}+\ket{\underbar{b}})(\bra{\bar{b}}+\bra{\underbar{b}})\ket{\psi_{\mathrm{ex}}^k}=0$. Therefore, $\ket{\psi_{\mathrm{ex}}^k}$ is obviously a ground state of $H_{\mathrm{ex}}^{\mathrm{DS}}$ with the lowest energy $0$. Next, we show that $H_{\mathrm{ex}}^{\mathrm{DS}}$ is gapped and $\ket{\psi_{\mathrm{ex}}^k}$ is unique in each topological sector. To this end, we define a \emph{non-local} unitary operator $\mathcal{V}$ such that $\mathcal{V}\ket{c}=-1^{N(\ket{c}
\star\ket{c_r})}\ket{c}$ for all $\ket{c}\in\mathcal{C}_0$. By calculating matrix elements of $\mathcal{V}H_{\mathrm{ex}}^{\mathrm{DS}}\mathcal{V}^\dagger$, it is easy to  show that $\mathcal{V}H_{\mathrm{ex}}^{\mathrm{DS}}\mathcal{V}^\dagger=t^2/V\sum_b(\ket{\bar{b}}-\ket{\underbar{b}})(\bra{\bar{b}}-\bra{\underbar{b}})$ which is the Hamiltonian $H_{\mathrm{ex}}$ for the TC topological order. Moreover, $\mathcal{V}\ket{\psi_{\mathrm{ex}}^k}=\sum_{c\in\mathcal{C}_0^k}\ket{c}$ is the ground state of $\mathcal{V}H_{\mathrm{ex}}^{\mathrm{DS}}\mathcal{V}^\dagger$ which has been shown to possess TC topological order. Therefore, according to the previous discussion, $\ket{\psi_{\mathrm{ex}}^k}$ is the unique ground state of $H_{\mathrm{ex}}^{\mathrm{DS}}$ in the sector $\mathcal{H}_0^k$ with an energy gap in the thermodynamic limit. In addition, using similar arguments as for the case of TC topological order, we can show that the ground states of $H_{\mathrm{ex}}^{\mathrm{DS}}$ are locally indistinguishable. Therefore, we conclude that $H_{\mathrm{ex}}^{\mathrm{DS}}$ is topologically ordered. By construction, $\mathcal{U}'H_{\mathrm{ex}}^{\mathrm{DS}}\mathcal{U}'^\dagger$ possess the same topological order as $H_{\mathrm{ex}}^{\mathrm{DS}}$. Finally, if the gap remains open when tuning $r$ to $0$, $H_{\mathrm{eff}}^{\mathrm{DS}}$ is also gapped and has the same ground-state degeneracy as $H_{\mathrm{ex}}^{\mathrm{DS}}$. As a result $H_{\mathrm{eff}}^{\mathrm{DS}}$ possesses the same topological order as $H_{\mathrm{ex}}^{\mathrm{DS}}$

In order to discuss anyonic excitations of $H_{\mathrm{ex}}^{\mathrm{DS}}$, we work with $\ket{\phi^k}=\mathcal{U}'\ket{\psi_{\mathrm{ex}}^k}=\sum_{c\in{\mathcal{C}_0^k}}-1^{N(\ket{c}\star\ket{c_r})}\ket{c}\star\ket{c_r}$ which is a sum of non-crossing-closed-loop configurations (see also Ref.~\cite{doublesemion1}). We can define the monomer, antimonomer, and vison excitations of $\ket{\psi_{\mathrm{ex}}^k}$ by directly generalizing the case where $\Lambda$ is the triangular lattice~\cite{doublesemion1}. Using the same arguments as in Ref.~\cite{doublesemion1} made for the case of triangular lattice, we can show that the three types of excitation define anyons of the DS topological order: semion, antisemion, and bosonic bound state. 

\section*{Discussion}
The most important conclusion that one can draw from the above discussion is that it is possible to realize the toric-code (TC) and the double-semion (DS) topological order in simpler, more realistic hamiltonians, specifically Bose-Hubbard-type models with two-site interaction terms. This becomes clear from the fact that ground states for the TC or DS topological order of Bose-Hubbard-type models discussed here can be locally unitarily transformed (and hence they possess the same topological order) into ground states which resemble ground-states of string-net models with the TC or DS topological order.
Indeed, the long-range entangled ground states for the TC or DS topological order of the models studied in this work can be locally unitarily transformed into a state $\ket{\psi}$. $\ket{\psi}$ is a sum of all Fock states in $\mathcal{C}_0$ which enter with equal-weight for the case of TC topological order or with a phase-factor for the case of DS topological order. By comparing $\ket{\psi}$ with the ground state of the exactly solvable models for the TC or DS topological order, one can see that their long-range-entanglement ``patterns'' are different. As discussed in Reference~\cite{topologicalorderbook1}, ``pattern'' of long-range entanglement refers to (1) common features describing Fock states (or spin product states) participating in the expansion of an entangled state, e.g., all Fock states in the expansion are closed-loop configuration; (2) how these Fock states participate in the expansion, i.e. their coefficients. Here, we discuss the TC topological order as an example. The DS topological order can be considered similarly. 

We define a generalized toric code on $\Lambda$, $H^{TC}=-\sum_{v}(-1)^{P(v)}A_{v}-\sum_{p}B_{p}$, where $A_v=\prod_{v\in\partial i}\sigma^z_i$ is defined on each vertex $v$, $B_p=\prod_{i\in\partial p}\sigma^x_i$ is defined on each plaquette $p$. Here, $\partial i$ denotes the two vertices delimiting link $i$, $\partial p$ is the set of links belonging to the boundary of plaquette $p$. $P(v)$ is the parity associated to vertices, as defined in the Topological sectors section. (Recall that spin $\downarrow$ and $\uparrow$ are identified with occupation numbers $1$ and $0$.)
By a local unitary transformation discussed in the Methods section, the model can be mapped to the original toric code~\cite{tc1}, so that $H^{TC}$ possesses the toric-code topological order. Moreover, the ground state $\ket{\Psi}$ of $H^{TC}$ is an equal-weight sum of \emph{all} Fock states in $\mathcal{C}$ (that is, all states which satisfy the fixed-parity condition given by $P(v)$). Due to the nature of the two-site density-density interaction and the fixed filling factor, $\mathcal{C}_0$ is a subset of $\mathcal{C}$. In other words, the sum defining $\ket\psi$ runs through only a part of states contributing to the sum defining $\ket\Psi$. By interpreting the equal-weight sum as long-range-entanglement ``pattern'', we conclude that $\ket{\psi}$ has a more restricted ``pattern'' than $\ket{\Psi}$, though both $\ket{\psi}$ and $\ket{\Psi}$ correspond to the same topological order. 

This difference in the long-range-entanglement ``pattern'' corresponds to different ways of realizing topological order. In a string-net model, the topological order is determined by local rules which determine spin product states expanding $\ket{\Psi}$ and the dynamics~\cite{sn1}. In a strong-interaction hardcore Bose-Hubbard-type lattice model, the topological order is determined by the ``local constraints'' (associated to density-density interaction and filling factor) which characterize Fock states expanding $\ket{\psi}$ and specify the dynamics. In light of the above discussion, we conjecture that there exists a correspondence between topological order realized in Bose-Hubbard-type lattice models and in string-net models. That is, given the long-range-entanglement ``pattern'' in ground states of certain string-net models, we can argue the existence of the same topological order in strong-interaction Bose-Hubbard-type lattice models by realizing a ``restricted pattern'' of long-range-entanglement in ground states.

\section*{Conclusions}
We have studied the emergence of the $\mathbb{Z}_2$ toric-code topological order and the double-semion topological order in a wide class of strongly-interacting hardcore lattice bosons described by Bose-Hubbard-type models Eq.~\ref{hamiltonian1} with density-density interaction. The model is defined on a non-bipartite lattice $\Lambda$ where degrees of freedom are located on links rather than sites. We have shown that model Eq.~\ref{hamiltonian1} harbors toric-code or double-semion topological order, with phase factor in the dynamics specified by Eq.~\ref{TC} or \ref{DS} respectively, if the following two conditions are satisfied. 

The first condition concerns ``local constraints'' determined by the interaction $H_0$ together with the fractional filling. ``Local constraints'', which characterize the Fock states spanning the lowest-energy subspace $\mathcal{H}_0$ of the interaction $H_0$, need to include the following two requirements. (i) The parity $P(v)$ assigned to each vertex $v$ is the same for all Fock states in $\mathcal{H}_0$. Here, $P(v)$ refers to the parity of $s(v)$, the number of links with occupation number $1$ adjacent to the vertex $v$. (ii) Positions of bosons in a given Fock state in $\mathcal{H}_0$ can be moved thus generating another Fock state in $\mathcal{H}_0$ in a way which generalizes how dimer configurations can be connected by resonances in quantum dimer models, as described in the Ergodicity section.

The second condition concerns the existence of an energy gap at $r=t^2/V$ of the $r$-parametrized hamiltonian $\tilde{H}(r)$ (given by Eq.~\ref{rparatc} and \ref{rparatcds} for the toric-code and the double-semion topological order, respectively) which connects an exactly solvable model to the effective hamiltonian. The gap needs to remain open when the parameter $r$ is tuned to $0$. 


We have shown that these two conditions are both determined by $H_0$ along with the fixed filling factor. Hence, these two conditions establish a connection between the density-density interaction and the emergence of the toric-code and the double-semion topological order of the hardcore lattice bosons in the strong-interaction limit.

We also conjectured that there exists a correspondence between hardcore Bose-Hubbard-type lattice models and string-net models by discussing the long-range-entanglement ``patterns''. That is, given the long-range-entanglement ``pattern'' in ground states of certain string-net models, we argued the existence of the {\em {same}} topological order in strong-interaction Bose-Hubbard-type lattice models through the realization of a ``restricted pattern'' of long-range-entanglement.


We expect that the work developed here will provide guidance to numerical studies which can pave the way for searching experimentally realizable lattice systems harboring topological order.

\section*{Methods}

\subsection*{Partition of $\mathcal{Z}$ and $\mathcal{C}$}
The formal summation ``$\star$'' defines an abelian group structure on the set of all Fock states. In this abelian group, $\mathcal{Z}$ is a subgroup and $\mathcal{C}$ is a coset of $\mathcal{Z}$. By restricting closed loops to be boundaries of regions of $\mathcal{S}$, we can define a subgroup $\mathcal{B}$ of $\mathcal{Z}$. The quotient group $\mathcal{Z}/\mathcal{B}$ is isomorphic to the first homology group of $\mathcal{S}$ and has $2^{2g}$ elements~\cite{homology1}. These elements, which are cosets of $\mathcal{B}$, are exactly $\{\mathcal{Z}^1,\cdots,\mathcal{Z}^{2^{2g}}\}$. The sets $\{\mathcal{C}^1,\cdots,\mathcal{C}^{2^{2g}}\}$ are also cosets of $\mathcal{B}$.

\begin{figure}
\centering
\includegraphics[width=0.94\textwidth]{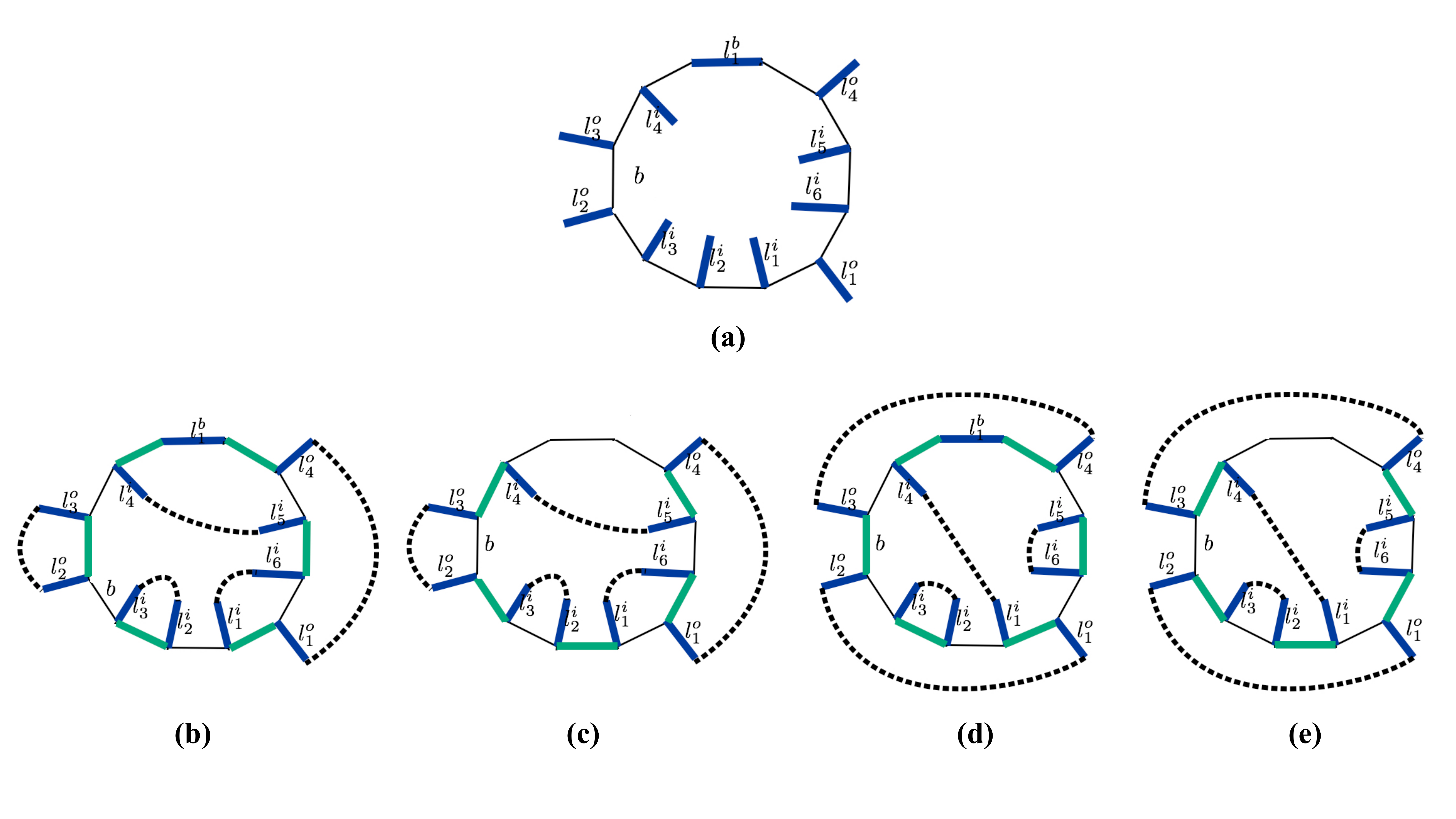}

\caption{\label{figfive}(\textbf{a}) represent an arbitrary loop $b$ contributing to the sum of $H_{\mathrm{eff}}^{\mathrm{DS}}$, where blue links emanating from $b$ and indexed clockwise by $l^o$, $l^i$ and $l^b$ represent occupied links in the reference state $\ket{c_r}$ while black links are unoccupied. We omit the lattice structure complementary to the loop $b$ and the links emanating from it. (\textbf{b}) and (\textbf{c}) represent the configuration of $\ket{c}\star\ket{c_r}$ and $\ket{c'}\star\ket{c_r}$, respectively. Blue links represent occupied links in $\ket{c_r}$, green links represent occupied links in $\ket{c}$ or $\ket{c'}$, black links are unoccupied, and black dashed lines complete loops outside and inside $b$. (\textbf{d}) and (\textbf{e}) represent the configuration of $\ket{c_1}\star\ket{c_r}$ and $\ket{c'_1}\star\ket{c_r}$, respectively. The significance of the colors is the same as above. Notice that loops outside and inside $b$ are connected differently than in (\textbf{b}) and (\textbf{c}).}
\end{figure}

\subsection*{\label{qb}Specifying $q(b)$ for the case of DS topological order}
In this section, we specify the phase factor $q(b)$ in Eq.~\ref{HexDS} for the DS topological order. Let's consider a contractible closed loop $b$ on $\Lambda$ which contributes to the sum in Eq.~\ref{HexDS}, a reference state $\ket{c_r}$ in $\mathcal{C}_0$, and an arbitrary state $\ket{c}$ in $\mathcal{C}_0$ such that bosons are located alternatively on $b$.  In the following, we will define $q(b)$ in terms of $b$ and $\ket{c_r}$ and show that, given the state $\ket{c'}=(\dyad*{\bar{b}}{\underbar{b}}+\dyad*{\underbar{b}}{\bar{b}})\ket{c}$, if $q(b)=1$ then the number of loops in $\ket{c'}\star\ket{c_r}$ has the same parity as in $\ket{c}\star\ket{c_r}$; if, instead, $q(b)=-1$ the parity is changed. Below, we illustrate arguments using figures, though we have formalized them into rigorous proof in the context of topological graph theory~\cite{topologicalgraphtheory} (not presented here for the sake of readability).

We start by discussing the relationship between the parity change under the action of $\dyad*{\bar{b}}{\underbar{b}}+\dyad*{\underbar{b}}{\bar{b}}$ on $\ket{c}$ and the links emanating from $b$ which, in the reference state $\ket{c_r}$, are occupied by bosons. Fig.~\ref{figfive}(a) shows state $\ket{c_r}$ on loop $b$, where thick blue links are occupied by bosons and black links are unoccupied (we omit the lattice structure complementary to $b$). We denote by $l^o$ the outer links emanating from $b$ with occupation number $c_r(l^o)=1$, $l^i$ the inner links emanating from $b$ with occupation number $c_r(l^i)=1$, and $l^b$ the links forming the loop $b$ with occupation number $c_r(l^b)=1$ (see Fig.~\ref{figfive}(a)).
Due to the fixed-parity condition, i.e. $P(v)=odd$ and $s(v)=1$ for all vertices, the number of $l^o$ links and the number of $l^i$ links are even. In order to show this, let's consider the non-crossing-closed-loop configurations of $\ket{c}\star\ket{c_r}$ or $\ket{c'}\star\ket{c_r}$ shown in Fig.~\ref{figfive}(b) and~\ref{figfive}(c) respectively, where blue links correspond to occupied links in $\ket{c_r}$, green links correspond to occupied links in $\ket{c}$ or $\ket{c'}$ respectively, black links are unoccupied links, and black dashed lines complete the loops outside and inside $b$. Note that, if a green and a blue link overlap, the link will appear black in the representation of $\ket{c}\star\ket{c_r}$ or $\ket{c'}\star\ket{c_r}$ due to the sum rule $1\star1=0$. Any string inside $b$ must start and end with $l^i$ links since no inner link emanating from $b$ has occupation number $1$ in $\ket{c}$ or $\ket{c'}$ (otherwise bosons cannot be located alternatively on $b$ in state $\ket{c}$ or $\ket{c'}$). Therefore, the total number of $l^i$ links is even. Similarly, the number of $l^o$ links is also even.
Notice that in the illustrations Fig.~\ref{figfive}(b) and \ref{figfive}(c), under the action of $\dyad*{\bar{b}}{\underbar{b}}+\dyad*{\underbar{b}}{\bar{b}}$ on $\ket{c}$, non-crossing closed loops are reconnected on loop $b$. In particular, as will be shown below, the reconnection is determined by how many inner links $l^i$  are in between two successive outer links $l^o$.

\begin{figure}
\centering
\includegraphics[width=0.94\textwidth]{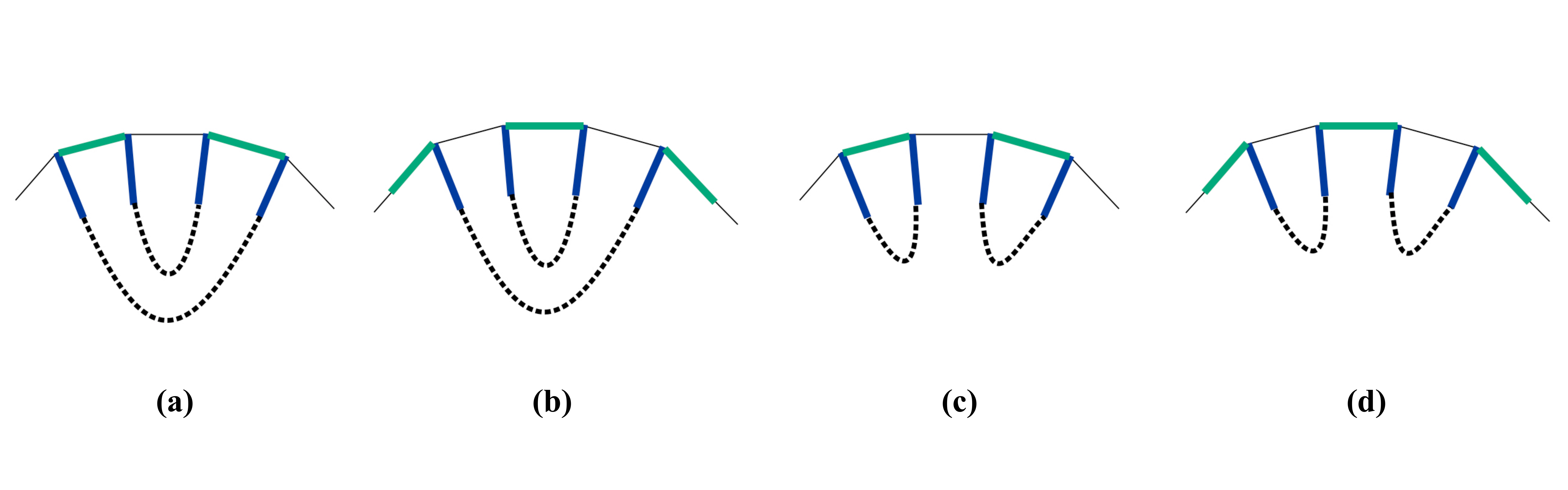}


\caption{\label{figsix}(\textbf{a}) and (\textbf{c}) illustrate an elementary loop reconnection operation which is also illustrated in (\textbf{b}) and (\textbf{d}). The loop reconnection does not affect the parity change upon shifting green links clockwise (or counterclockwise) on $b$. Colored links have the same meaning as in previous figures.}
\end{figure}

We now show that the parity change under the action of $\dyad*{\bar{b}}{\underbar{b}}+\dyad*{\underbar{b}}{\bar{b}}$ is independent of $\ket{c}$. For this purpose, let us consider another state $\ket{c_1}$ in $\mathcal{C}_0$, in which bosons are located alternatively on $b$. Let $\ket{c'_1}=(\dyad*{\bar{b}}{\underbar{b}}+\dyad*{\underbar{b}}{\bar{b}})\ket{c_1}$. We set $\ket{c_1}$  to coincide with $\ket{c}$ on $b$ and $\ket{c'_1}$ to coincide with $\ket{c'}$ on $b$. The configurations of $\ket{c_1}\star\ket{c_r}$ and $\ket{c'_1}\star\ket{c_r}$ are represented by Fig.~\ref{figfive}(d) and \ref{figfive}(e) respectively. Notice that loops in $\ket{c_1}\star\ket{c_r}$ and $\ket{c'_1}\star\ket{c_r}$ are connected in a different way from those in $\ket{c}\star\ket{c_r}$ and $\ket{c'}\star\ket{c_r}$. We can show that, if, upon reconnecting loops going from Fig.~\ref{figfive}(b) to \ref{figfive}(d), the parity of the number of loops is unchanged (changed), then, upon reconnecting loops going from Fig.~\ref{figfive}(c) to \ref{figfive}(e), the parity of the number of loops is also unchanged (changed). This can be shown from an elementary loop reconnection illustrated in Fig.~\ref{figsix}, where the colored links are interpreted in the same way as in previous figures. Note that Fig.~\ref{figsix}(a) coincides with \ref{figsix}(c) on $b$ though loops are reconnected in such a way that \ref{figsix}(c) possesses one extra loop. Likewise, \ref{figsix}(b) coincides with \ref{figsix}(d) on $b$ but \ref{figsix}(d) possesses one less loop than \ref{figsix}(b). 
Therefore, upon shifting the green links clockwise (or counterclockwise), which is equivalent to acting with $(\dyad*{\bar{b}}{\underbar{b}}+\dyad*{\underbar{b}}{\bar{b}})$ on $\ket{c}$, the parity change from Fig.~\ref{figsix}(a) to \ref{figsix}(b) is the same as from Fig.~\ref{figsix}(c) to \ref{figsix}(d). As a result, since the action of $\dyad*{\bar{b}}{\underbar{b}}+\dyad*{\underbar{b}}{\bar{b}}$ only reconnects loops on $b$, the parity change in the number of loops in going from $\ket{c}\star\ket{c_r}$ to $\ket{c'}\star\ket{c_r}$ is the same as the parity change in the number of loops in going from $\ket{c_1}\star\ket{c_r}$ to $\ket{c'_1}\star\ket{c_r}$, as was shown in Fig.~\ref{figfive}. In other words, the parity change is not affected by how links are connected outside or inside the loop $b$. Simply put it, dashed lines in Fig.~\ref{figfive} do not matter.

For simplicity, in the following, we equivalently work on the surface $\mathcal{S}$ on which the lattice $\Lambda$ is embedded and interpret Fig.~\ref{figfive}(b), \ref{figfive}(c), \ref{figfive}(d) and \ref{figfive}(e) as the images of links and loops on $\mathcal{S}$. 
With this interpretation, we are allowed to modify loop $b$ by removing links without worrying about the lattice structure as long as green links are alternating on loop $b$ and each site on $b$ possesses exactly one blue link. As we explain below in details, this procedure allows us to determine the parity change in question. 
To this end, we define two even numbers $M_1(b,c_r)$ and $M_2(b,c_r)$ as follows.

\begin{figure}
\centering
\includegraphics[width=0.94\textwidth]{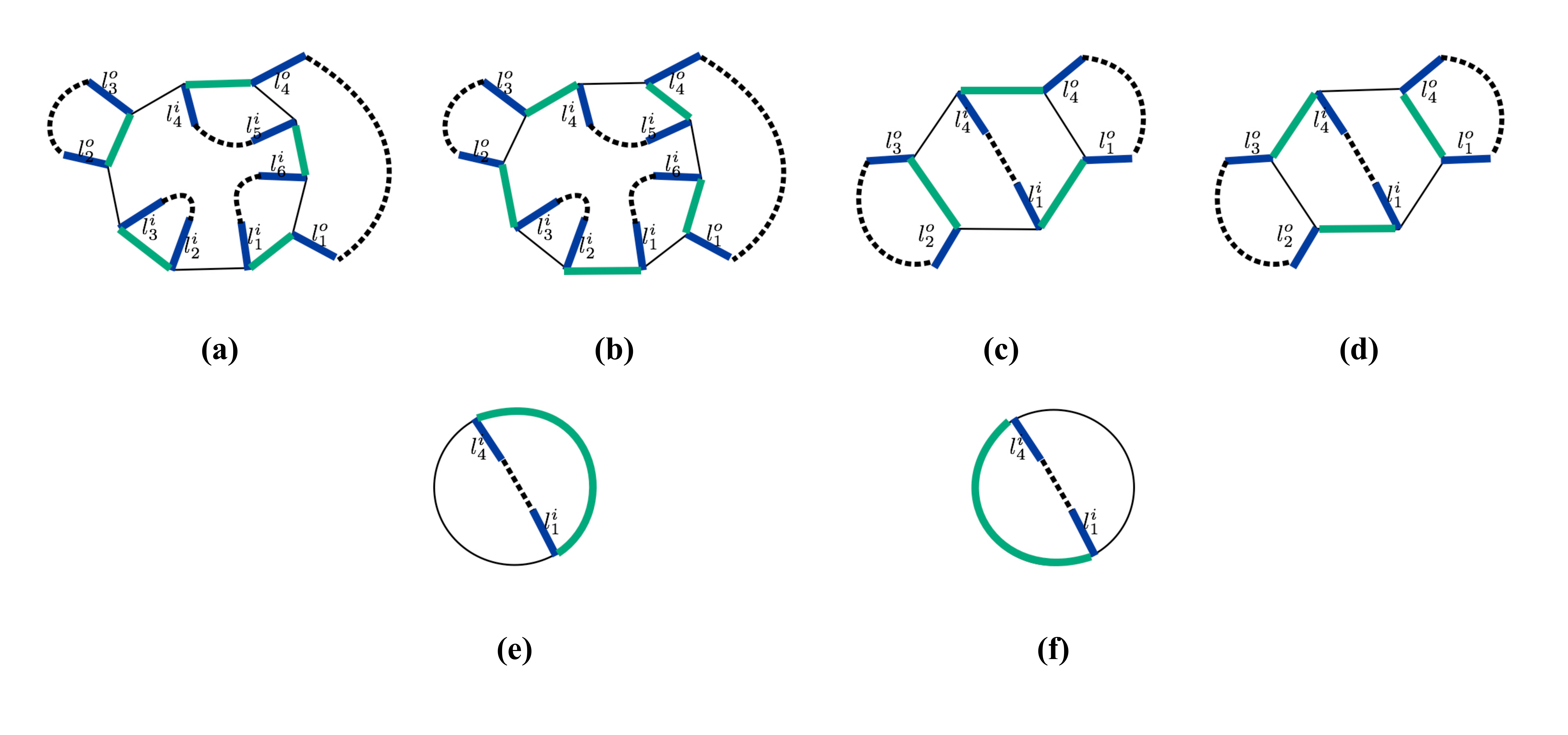}


\caption{\label{figseven}(\textbf{a}) and (\textbf{b}) show configurations obtained after removing $l^b$ links from Fig.~\ref{figfive}(b) and \ref{figfive}(c). Note that the parity change is the same as before removing $l^b$ links. (\textbf{c}) and (\textbf{d}) show configurations obtained after removing links $l^i_2$, $l^i_3$, $l^i_5$ and $l^i_6$ from (\textbf{a}) and (\textbf{b}). Note that, since an even number of pairs of links has been removed, the parity change stays the same. (\textbf{e}) and (\textbf{f}) show configurations obtained after removing links $l^o_1$, $l^o_2$, $l^o_3$ and $l^o_4$ from (\textbf{c}) and (\textbf{d}). Again, since an even number of pairs of links has been removed, the parity change stays the same. Colored links and dashed links have the same meaning as in previous figures.}
\end{figure}

We start by removing $l^b$ links. As shown in Fig.~\ref{figfive}(b) and \ref{figfive}(c), link $l^b_1$ either forms a segment of a closed loop together with green links, or overlaps with a green link and does not contribute to any loop. As a consequence, removing $l^b$ links has no effect on the number of loops and hence on the parity change. Fig.~\ref{figseven}(a) and \ref{figseven}(b) show the new configurations obtained after removing $l^b$ links from Fig.~\ref{figfive}(b) and \ref{figfive}(c). Notice that, in order for the number of links on $b$ to stay even, links must be removed in pairs.

We now fix any two successive $l^o$ links and remove an even number of $l^i$ links existing in between until there is at most one such $l^i$ link.  As an example of such removal, we remove links $l^i_2$, $l^i_3$ between links $l^o_1$ and $l^o_2$, and $l^i_5$, $l^i_6$ between links $l^o_1$ and $l^o_4$ from Fig.~\ref{figseven}(a) and \ref{figseven}(b). Note that, for each $l^i$ removed, an adjacent link on $b$ needs to be removed in order to have exactly one blue link attached to each vertex. After the removal, we obtain configurations Fig.~\ref{figseven}(c) and \ref{figseven}(d). The parity change in the configurations obtained after this removal is the same as before if we remove an even total number of pairs while the parity change is opposite if the total number of removed pairs is odd. To see this, consider such a pair of $l^i$ links, say $l^i_2$ and $l^i_3$ in Fig.~\ref{figseven}(a) and \ref{figseven}(b), which are connected by a black dashed line. Note that, if such a pair is not connected by a black dashed line, we can always reconnect loops so that the pair is connected by a black dashed line inside $b$ without affecting the parity change. $l^i_2$ and $l^i_3$ form a loop together with a green link and a black dashed line in Fig.~\ref{figseven}(a). After shifting green links clockwise on $b$, this loop no longer exists in Fig.~\ref{figseven}(b). As a result, the removal of an odd number of pairs of $l^i$ links results in the opposite parity change compared to the parity change of the initial configurations.

Subsequently, in the same manner, we remove an even number of $l^o$ links which are located in between two consecutive $l^i$ links. Once again, the removal alters the parity change in the same way as explained above for removal of $l^i$ links.

Finally, we repeat these two removal processes until meeting one of these three situations: (1) there are only $l^o$ links; (2) there are only $l^i$ links; (3) $l^o$ links and $l^i$ links are located alternatively along the loop $b$. As an example, we further remove links $l^o_1$, $l^o_2$, $l^o_3$ and $l^o_4$ from Fig.~\ref{figseven}(c) and \ref{figseven}(d) to obtain the final configurations Fig.~\ref{figseven}(e) to \ref{figseven}(f).

\begin{figure}
\centering
\includegraphics[width=0.94\textwidth]{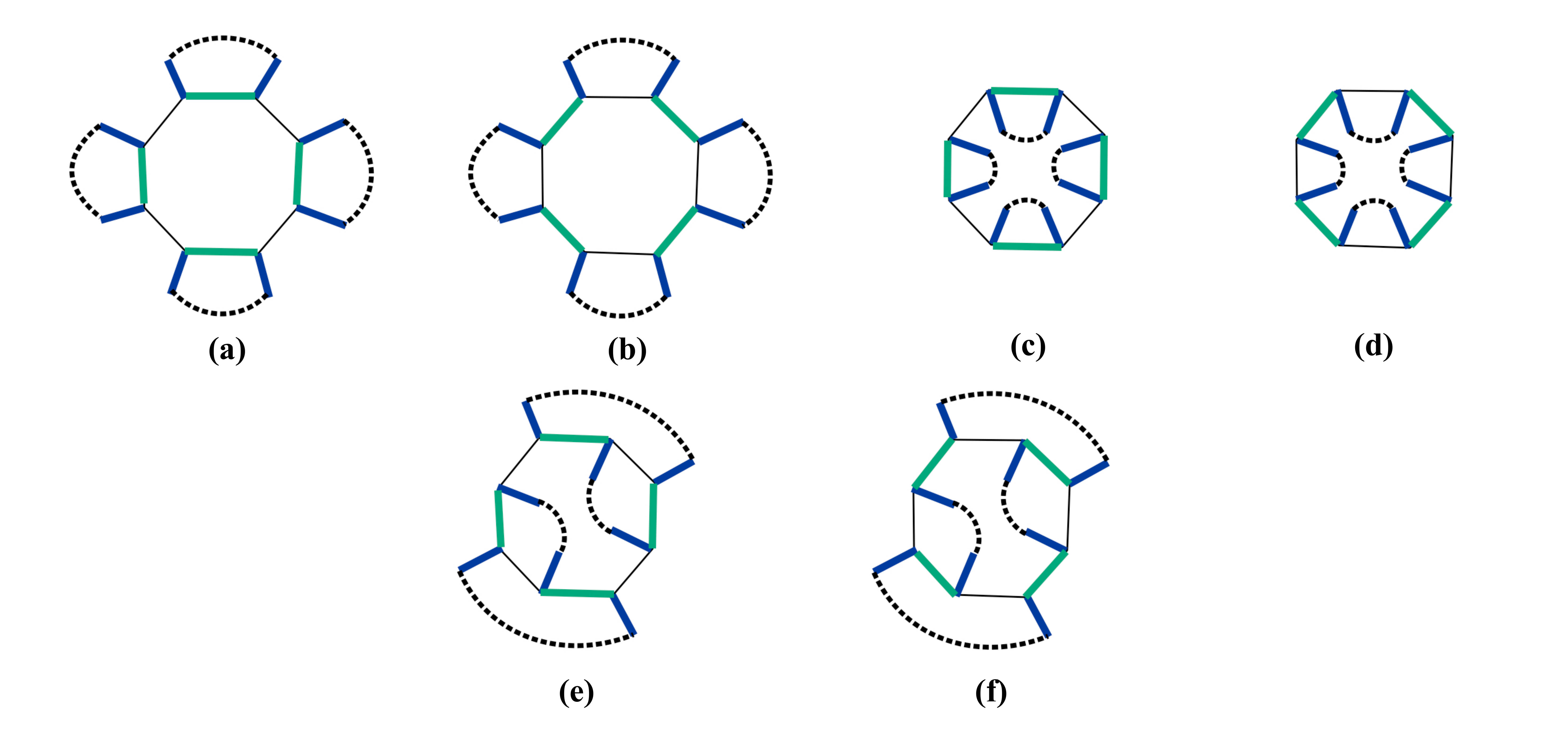}


\caption{\label{figeight}(\textbf{a}) and (\textbf{b}) show parity change in situation when there are only $l^o$ links; (\textbf{c}) and (\textbf{d}) show parity change in situation when there are only $l^i$ links; (\textbf{e}) and (\textbf{f}) show parity change in situation when $l^o$ and $l^i$ links are located alternatively along loop $b$. Colored links and dashed lines have the same meaning as in previous figures.}
\end{figure}

We are now ready to define $M_1(b,c_r)$ and $M_2(b,c_r)$ as follows. $M_1(b,c_r)$ is the total number of $l^o$ and $l^i$ links removed. $M_2(b,c_r)$ is the number of $l^o$ links if we end up in situation (1), the number of $l^i$ links if we end up in situation (2), or the number of $l^o$ (or $l^i$) links if we end up in situation (3). Note that $M_1(b,c_r)$ is even by definition and $M_2(b,c_r)$ is even since the number of $l^o$ links and the number of $l^i$ links are both even.

Let us consider the example shown in Fig.~\ref{figeight}, where \ref{figeight}(a) and \ref{figeight}(b) show the parity change in situation (1); \ref{figeight}(c) and \ref{figeight}(d) show the parity change in situation (2); and \ref{figeight}(e) and \ref{figeight}(f) show the parity changein situation (3). One can easily check that $M_2(b,c_r)$ as defined above characterizes the parity change as follows. If $i^{M_2(b,c_r)+2}=1$, then the parity change stays the same under the action of shifting green links clockwise, while if $i^{M_2(b,c_r)+2}=-1$, then the parity change is opposite. 

Furthermore, recall that $M_1(b,c_r)$, by definition, records the total number of $l^o$ and $l^i$ links removed and, as a consequence, keeps track of whether the parity change between initial configurations is the same or not as in the final configurations. Recall, an even number of removed pairs implies the same parity change, while an odd number of removed pairs implies the opposite parity change. Then, combining $M_1(b,c_r)$ and $M_2(b,c_r)$, we can conclude that $i^{M_1(b,c_r)+M_2(b,c_r)+2}$ gives the parity change between the initial configurations. 


\subsection*{Local unitary transformation connecting the generalized toric code and the original toric code}
We fix a reference Fock state $\ket{c_r}$ in $\mathcal{C}$ and define $\mathcal{U}'=\prod_{{c_r}_i=1}\sigma^x_i$. Obviously $\mathcal{U}'$ is a quantum circuit and thus local unitary. It is easy to check that $\mathcal{U}'H^{TC}\mathcal{U}'^\dagger=-\sum_{v}A_{v}-\sum_{p}B_{p}$ which is the original toric-code. Due to the unitarity, $\mathcal{U}'H^{TC}\mathcal{U}'^\dagger$ and $H^{TC}$ have the same spectrum. The ground state of $H^{TC}$ is $\sum_{\ket{z}\in\mathcal{Z}}\mathcal{U}'^\dagger\ket{z}=\sum_{\ket{z}}(\ket{z}\star\ket{c_r})=\sum_{\ket{c}\in\mathcal{C}}\ket{c}$

\subsection*{Data Availability}
No datasets were generated or analysed during the current study.

\bibliography{jul062017}

\section*{Acknowledgments}
We thank Mohammad Maghrebi, Arghavan Safavi-Naini, Guanyu Zhu, Fabio Lingua and Chao Zhang for useful discussion. This work is supported by NSF (PIF-1552978).

\section*{Author contributions}
W.W. did the analysis and calculation. B.C. coordinated the project. Both authors prepared the manuscript.

\section*{Additional Information}
\textbf{Competing financial interests:} The authors declare no competing financial interests.

\newpage

\end{document}